\documentclass[journal,twoside,web]{IEEEtran}
\usepackage{cite}
\usepackage{amsmath,amssymb,amsfonts}
\usepackage{algorithmic}
\usepackage{graphicx}
\usepackage{textcomp}
\usepackage{subcaption} 

\def\BibTeX{{\rm B\kern-.05em{\sc i\kern-.025em b}\kern-.08em
    T\kern-.1667em\lower.7ex\hbox{E}\kern-.125emX}}
\begin{document}

\title{Segmentation of Breast Microcalcifications: a Multi-Scale Approach}
\author{Chrysostomos Marasinou, Bo Li, Jeremy Paige, Akinyinka Omigbodun, Noor Nakhaei,\\ Anne Hoyt, William Hsu, Senior Member, IEEE
\thanks{
This work was supported in part by the National Science Foundation under Grant No. 1722516 and the Department of Radiological Sciences through the generous support of the Iris Cantor Foundation.}
\thanks{C. Marasinou, L. Bo, J. Paige, A. Omigbodun, A. Hoyt, and W. Hsu are with the David Geffen School of Medicine at UCLA, Department of Radiological Sciences, Los Angeles, CA 90024, United States of America (email: whsu@mednet.ucla.edu)}
\thanks{N. Nakhaei is with the University of California, Los Angeles, Department of Computer Science, Los Angeles, CA 90024, United States of America}
}

\maketitle

\begin{abstract}
Accurate characterization of microcalcifications (MCs) in 2D full-field digital screening mammography is a necessary step towards reducing diagnostic uncertainty associated with the callback of women with suspicious MCs. Quantitative analysis of MCs has the potential to better identify MCs that have a higher likelihood of corresponding to invasive cancer. However, automated identification and segmentation of MCs remains a challenging task with high false positive rates. We present Hessian Difference of Gaussians Regression (HDoGReg), a two stage multi-scale approach to MC segmentation. Candidate high optical density objects are first delineated using blob detection and Hessian analysis. A regression convolutional network, trained  to output a function with higher response near MCs, chooses the objects which constitute actual MCs. The method is trained and validated on 435 mammograms from two separate datasets. HDoGReg achieved a mean intersection over the union of 0.670$\pm$0.121 per image, intersection over the union per MC object of 0.607$\pm$0.250 and true positive rate of 0.744 at 0.4 false positive detections per $cm^2$. The results of HDoGReg perform better when compared to state-of-the-art MC segmentation and detection methods.
\end{abstract}

\begin{IEEEkeywords}
breast cancer, full-field digital mammography, microcalcifications, segmentation
\end{IEEEkeywords}

\section{Introduction}\label{sec:I}




Breast cancer is the most prevalent cancer among women. More than two million new cases and over 620,000 deaths were estimated for 2018 worldwide \cite{Bray2018}.  Studies have shown that early detection using mammography reduces breast cancer mortality \cite{Marmot2013}. In many countries, screening programs have been established with sensitivity levels for detecting cancers ranging between 80-95\%  \cite{doi:10.1002/ijc.29593, Euler-Chelpin2018}. However, screening also results in false positive outcomes, including the identification of clinically insignificant cancers, raising concerns about overdetection.

Microcalcifications (MCs), which are small calcium deposits, are common mammographic findings where they typically appear as bright structures. Nearly 50\% of the biopsied MCs are associated with ductal carcinoma in situ (DCIS) \cite{Farshid2011}, a largely benign form of cancer but a nonobligate precursor to invasive cancer \cite{Hofvind2011, Cox2012}. MCs are reported by radiologists using a set of qualitative descriptors related to shape (morphology) and distribution, as defined by the American College of Radiology Breast Imaging Reporting and Data System (ACR BI-RADS). Descriptors correspond to varying levels of suspicion for cancer. For example, amorphous MCs are assigned a moderate suspicion level (i.e., BI-RADS 4B) with a positive predictive value (PPV) of 21\%  \cite{ACR_BI-RADS_5th}. However, interpretation of MCs varies by radiologist due to subtle differences in their size, shape, texture, and inhomogeneity of the background tissue \cite{Wang2018}. Hence, determining whether a group of calcifications is malignant is challenging, and the current PPV of assessing suspicious MCs on 2D mammography is in the range of 20-41\% \cite{Wilkinson2017}.

Many computerized methods have been developed to aid radiologists in detecting MCs \cite{Dengler1993,Yoshida1994,Netsch1999,El-Naqa2002,Oliver2012, Wang2018}. These methods, generally categorized as computer-aided detection (CADe) systems, automatically mark groups of suspicious MCs in mammograms. Current CADe systems achieve high sensitivity but at the cost of a large number of false positive marks per mammogram, increasing the interpretation time. Furthermore, most of the automated methods identify regions with suspicious MCs but do not delineate the exact shape of individual MCs. Studies have shown that using shape and intensity features from segmented MCs can improve malignancy classification \cite{Strange2014, Bria2014, Chen2015, Alam2019, Cai2019}. Precise segmentation also allows for more accurate quantitative characterization of shape and distribution of MCs and texture analysis of the surrounding breast parenchyma that could be used to classify cancerous regions better.


In this paper, a quantitative approach for characterizing MCs based on their shape is demonstrated. Given a 2D full-field digital mammogram, we initially identify bright salient structures using the difference of Gaussians’ (DoG) blob detection algorithm. Hessian analysis is then applied to segment these structures. Next, dense regression is applied to segment regions containing structures that are likely to be MCs. Dense regression has been used for similar tasks such as cell \& nuclei detection \cite{10.1007/978-3-319-24574-4_33, XIE2018245}, retinal optical disc and fovea detection \cite{Meyer2018}, and focal vascular lesion localization on brain MRI \cite{10.1007/978-3-030-32251-9_26}. The idea is that human experts' reference annotations are mapped to a smooth proximity function that reaches its maximum value when corresponding to the annotated points. Dense regression models are then trained to map the input mammogram to the proximity function. The proximity function method is advantageous when the manual annotations are incomplete (e.g., impractical to annotate every possible MC) or when objects are annotated by a single pixel rather than their actual boundaries (e.g., many MCs are tiny and time-consuming to delineate the boundary).  To perform dense regression, a pretrained, fully convolutional network with pretrained weights is utilized. The outputs of the dense regression model and the blob segmentation algorithm are combined to generate the final MC segmentation. 

The novelty of our work is summarized as follows:
\begin{itemize}
    \item  Proximity functions represent individual MCs, providing a way to accommodate uncertainty in boundaries as part of classifier training.
    \item  A dense regression model and a novel blob segmentation algorithm are applied to generate MCs' accurate segmentation while achieving fewer false positives than other state-of-the-art algorithms. 
\end{itemize}

\subsection{Prior Work}
Detection of MCs generally occurs in two steps: (i) detection of individual MCs, (ii) identification of clusters by grouping individual MCs following a set of spatial clustering rules. For detecting individual MCs, various approaches have been investigated, ranging from traditional image processing techniques to deep learning methods. \cite{Dengler1993} used a DoG filter for spot detection and shape reconstruction by morphological filtering. \cite{Yoshida1994} used 2D wavelet transform to suppress the low frequency background and enhance small signals. \cite{Netsch1999} applied a multiscale spot detector based on Laplacian filters and size and contrast estimation based on filter response. \cite{El-Naqa2002} used support vector machines with the input being local image windows to classify locations as MCs or non-MCs. \cite{Oliver2012} extracted local features using a predefined filter bank and applied supervised learning with a boosting classifier. Most of these methods focused on detecting the point locations of MCs, disregarding the task of extracting their shape. More recently, \cite{Wang2018} developed a context-sensitive deep convolutional neural network to classify MCs' presence at locations. In their approach, candidate locations were generated using DoG blob detection, filtering out non-MCs using a convolutional neural network. In contrast, our study performs MC segmentation (i.e., each MC is characterized by its shape and spatial location). 

Several studies have examined MC segmentation. Techniques include wavelet transform to isolate high frequency components \cite{Strickland2002, Regentova2007}, gray-level morphological operations \cite{ Dengler1993, Betal1997, Halkiotis2002, Xu2011, Ciecholewski2017}, fuzzy logic \cite{Heng-DaCheng2002}, and machine learning for binary pixel classification \cite{Bria2014}. Although segmentation of MCs is performed in these studies, all but one \cite{Ciecholewski2017}) evaluated the performance of their algorithm as a detection task, not a segmentation task. Reference \cite{Ciecholewski2017} reported an intersection over the union of 70.8\% between the segmented MCs and the radiologist annotations on a set of 200 regions. We have reimplemented their approach, comparing its performance to our HDoGReg approach using the same dataset.


\subsection{Paper Organization}
The rest of the paper is organized as follows. In section \ref{sec:II}, we present the data utilized for training and validating our model. In section \ref{sec:III}, we formulate the problem and describe the methodology of Hessian Difference of Gaussians Regression (HDoGReg) and the corresponding evaluation metrics. The experiments for validating our models are described in section \ref{sec:IV}, where the comparison methods are given. In section \ref{sec:V}, we present and discuss the results. Finally, conclusions are drawn in section \ref{sec:VI}.

\section{Data}\label{sec:II}

Two distinct full-field digital mammography datasets were used in this work: The publicly available INbreast dataset and cases seen at our institution (UCLA).

\textit{INbreast}: For model training and internal validation, we utilized a public dataset called INbreast \cite{MOREIRA2012236}, a collection of 2D mammograms generated using a Siemens MammoNovation system. A total of 115 screening cases with 410 images were collected at a 0.070 mm per pixel resolution and 14-bit greyscale. The dataset included detailed annotations provided by two experts for several types of lesions (i.e., masses, MCs, asymmetries, and distortions). Fifty-six cases had pathology-confirmed diagnoses, out of which 45 were cancerous (DCIS and invasive). We used 294 images with annotations of individual MCs. MCs were annotated in two ways: (1) small MCs were annotated by a single pixel to denote their location and (2) larger MCs were annotated using pixel-wise contours. It should be noted that the guideline of what was considered small versus larger MC was not reported \cite{MOREIRA2012236}.

\textit{UCLA}: As an additional external test set, we utilized data collected retrospectively from patients who had a mammogram performed by our institution, following an institutional review board-approved protocol. The dataset consisted of 79 diagnostic cases with 141 full-field digital images where MCs were present. All images were acquired using Hologic Selenia full-field digital mammography devices at a 0.070 mm per pixel resolution and 12-bit greyscale. After collecting the data, suspicious MCs were annotated by a breast fellowship-trained, board-certified radiologist. An open-source medical image viewer, Horos, was utilized to generate the annotations. Individual MCs were annotated by single pixels indicating their locations. To assess the annotation task's interreader reliability, a sample of 5 cases was annotated by a second board-certified radiologist. The index of specific agreement and the kappa statistic were determined. The two radiologists' agreement was moderate, with an index of specific agreement of 0.664 (0.606-0.729, 95\% confidence intervals), see supplementary material\footnote{Supplementary material will be made available when published}.

\section{Methods}\label{sec:III}
\begin{figure}
    \centering
    \includegraphics[width=\linewidth]{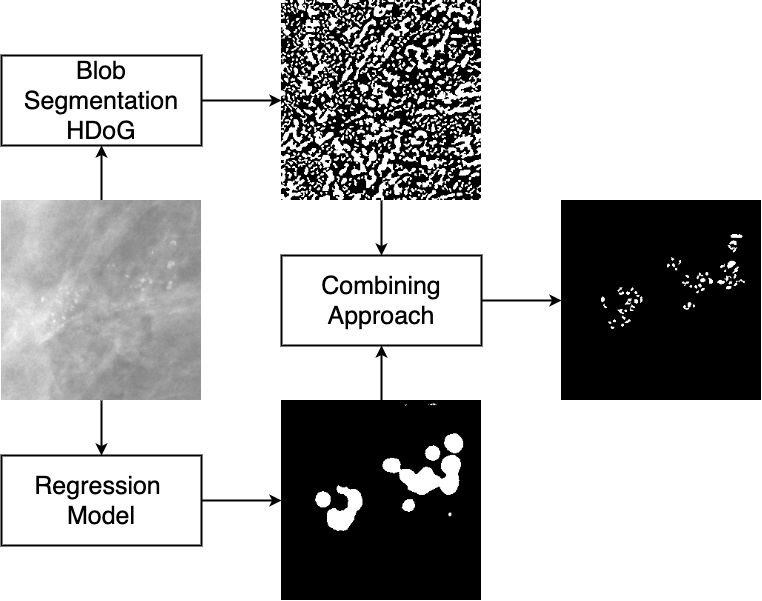}
    \caption{HDoGReg: Approach for segmenting individual MCs. While the segmentation is performed on entire 2D full-field mammograms, for visualization purposes, a small patch is shown. In the upper branch, blob segmentation is performed to segment bright blob-like and tubular structures. In the lower branch, a regression convolutional neural network gives a continuous function with higher response close to MCs. A threshold is then applied to segment regions where MCs are likely present. The two branches' output is combined based on an overlap criterion, resulting in the final segmentation mask.}
    \label{fig:SegmentationPipeline}
\end{figure}

The overall approach to MC segmentation, HDoGReg, is illustrated in Figure \ref{fig:SegmentationPipeline}. 

\subsection{Blob segmentation}
The first stage in HDoGReg is the segmentation of granular structures that are candidate MCs. To generate candidate MC segments, we developed Hessian DoG for blob segmentation. This module's objective is to obtain an accurate segmentation of bright salient structures that are candidate MC objects, as shown in Figure \ref{fig:SegmentationPipeline}.

Scale-space theory is a framework formulated to represent signals at multiple scales. The Gaussian scale-space representation of an image $I(x,y)$ is defined as \cite{Lindeberg1998}:
\begin{align}
L(x, y; \sigma) = G(x, y; \sigma) \ast I(x, y),
\end{align}
where $\ast$ is the convolution and $G(x,y,\sigma)$ the two-dimensional Gaussian function
\begin{align}
G(x,y; \sigma) &= \frac{1}{2\pi \sigma^2} e^{-(x^2+y^2)/(2\sigma^2)}
\end{align}
In the DoG method, blobs with associated scale levels are detected from scale-space maxima of the scale-normalized DoG function. The normalized DoG function is defined as:
\begin{align}\label{eq:DoG}
DoG(x,y;\sigma) &= \frac{\sigma}{\Delta \sigma} 
\left(
L(x,y;\sigma+\Delta\sigma)-L(x,y;\sigma)
\right)
\end{align}
where $\Delta \sigma$ is the difference between two scales. To construct the DoG scale-space representation, a sequence of scales is considered $\sigma_n = k^n \sigma_\textrm{min}$ where $k$ is a constant multiplicative factor and $n=[0,1,\cdots, n_\textrm{max} ]$. The DoG representations \eqref{eq:DoG} are computed for all adjacent scales (i.e., $\Delta \sigma = \sigma_{n+1}-\sigma_n$) forming a 3-dimensional representation:
\begin{align}
DoG(x,y,n) &= \frac{\sigma_n}{\sigma_{n+1}-\sigma_n} 
\left(
L(x,y;\sigma_{n+1})-L(x,y;\sigma_n)
\right)
\end{align}
with $x,y$ the two spatial dimensions and $n=[0,1,\cdots, n_{\max-1} ]$ a scale dimension. Local-maxima in the 3-dimensional representation are computed giving a blob set $(x^{(i)},y^{(i)},\sigma^{(i)})$ where $i$ identifies each blob. The number of blob detections is controlled by a threshold, $T_\textrm{DoG}$, that is applied as a lower bound on the stacked representation before obtaining the local maxima. 
Moreover, in the case of overlapping blobs, the smaller blob is eliminated if the overlapping fraction is greater than the threshold $O_\textrm{DoG}$.

We extended this method using Hessian analysis to achieve blob segmentation. The geometrical structure of a blob-like object can be described by the eigenvalues of the Hessian \cite{10.1007/BFb0056195}. In particular, a bright blob-like structure corresponds to two negative and large eigenvalues, whereas a bright tubular structure corresponds to one large negative eigenvalue and a small eigenvalue of an arbitrary sign. These structures correspond to the target MC candidates.

The Hessian of the DoG representation at scale $\sigma$ is given by

\begin{align}
\textrm{HDoG}(x,y;\sigma) =
\begin{pmatrix}
\frac{\partial^2 DoG(x,y;\sigma)}{\partial x^2} & \frac{\partial^2 DoG(x,y;\sigma)}{\partial x \partial y}\\
\frac{\partial^2 DoG(x,y;\sigma)}{\partial x \partial y} & \frac{\partial^2 DoG(x,y;\sigma)}{\partial y^2}
\end{pmatrix}
\end{align}
HDoG is computed across all scales in the sequence $\sigma_n$. At each scale the following constraints are imposed:
\begin{align}
\textrm{tr}(H) < 0 \quad \land \quad  \left(\det(H)<0\quad  \lor \quad \frac{|\det(H)|}{\textrm{tr}(H)^2} \leq h_{\textrm{thr}} \right)
\end{align}
where $h_{\textrm{thr}}$ is a tunable parameter. The constraints ensure that the Hessian is either negative definite or has a small positive eigenvalue. In this way, only bright salient blob-like and tubular structures are segmented. The constraint generates a binary mask at each scale. Iterating over the blob set found in the DoG algorithm $(x^{(i)},y^{(i)},\sigma^{(i)})$, the corresponding objects are found in the Hessian masks. More specifically, for the Hessian mask at scale $\sigma^{(i)}$, the object spanning the location  $(x^{(i)},y^{(i)})$ is found. The final blob segmentation result is comprised of all detected objects merged into a single binary mask.

Our HDoG algorithm works on a predetermined scale range $[\sigma_{\min} , \sigma_{\max}]$, which was fine-tuned on the validation set to
achieve a high true positive rate (\textgreater 90\%). The segmented objects from this phase were considered candidate MC objects and filtered in the subsequent step to reduce the number of false positives.

\subsection{Regression Convolutional Neural Network}

\begin{figure*}[!t]
    \centering
    \includegraphics[width=0.95\linewidth]{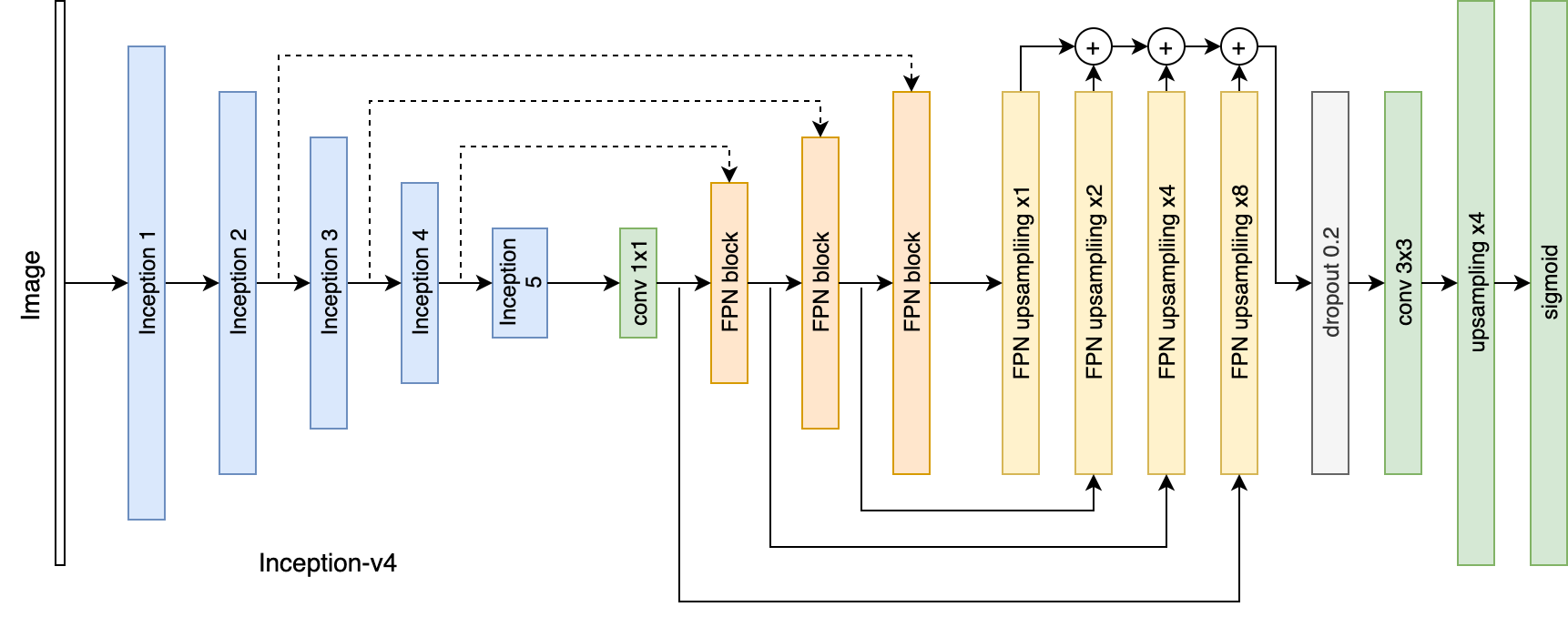}
    \caption{FPN with Inception-v4 encoder used for regression. In the encoding branch the image is processed with the Inception-v4 classification network. Skip connections (dashed lines) are inserted after layers where the output was reduced in spatial size by factors of 4, 8, 16 and 32 respectively. The skip connections feed FPN blocks where they are processed by convolutions. The outputs are up-sampled independently by factors of 1, 2, 4 and 8 respectively. Their outputs are added and inserted in a spatial dropout layer activated only during training for regularization purposes. After dropout, convolutions are followed by an up-sampling by a factor of 4 to match the original image size and the sigmoid activation function.}
    \label{fig:CNN}
\end{figure*}

In the second stage, we identified regions where MCs are most likely to be present in the mammogram. The task here was to segment the MCs' area and use the segmentation to choose relevant MC objects from the previous stage. We performed this task using a fully convolutional neural network, previously used in image segmentation, as a regression model. The model's output is a smooth proximity map reaching a maximum value at the predicted MC locations.

Segmentation approaches commonly perform binary classification on each pixel. However, applying this method requires precise pixel-wise annotations for all objects of interest appearing in our images. As previously described, generating precise annotations of MCs is challenging, and the reference annotations only provide the point locations of MCs, not their exact boundaries.

The task of MC region segmentation is analogous to that of cell \& nuclei detection in microscopy images. The two tasks share the following characteristics: (1) they are highly imbalanced ( i.e., the positive class [MCs or cells/nuclei] captures a small region compared to the background within an image, and it often consists of many small structures), (2) the background is highly inhomogeneous, (3) individual objects (MCs or cells/nuclei) exhibit large variation in sizes, shapes, and textures, (4) boundaries of the structures are often blurry, (5) the resolution of both types of images is large, and (6) the annotations are usually a mixture of individual points or exact boundaries.

Inspired by this analogy, we adapted methods previously used in cell \& nuclei detection \cite{10.1007/978-3-319-24574-4_33, XIE2018245}. In \cite{10.1007/978-3-319-24574-4_33}, the authors proposed regression as a method for detecting cell centers. The human-annotated binary masks containing cell centers' locations were transformed into a continuous function flat on the background with localized peaks at each cell's center. These functions were then used to train a Random Forest Regression algorithm on a set of image patches. The cell centers were identified with a local maxima in the model’s output. In \cite{XIE2018245}, the authors showed that the same technique could be applied using a deep learning model. Their regression model was a fully convolutional neural network with a large receptive field capable of encoding high-resolution information.

\begin{figure}
  \centering
  \begin{subfigure}{0.3\linewidth}
    \centering
    \includegraphics[width=\linewidth]{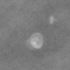}
    \caption{}
  \end{subfigure}
  \begin{subfigure}{0.3\linewidth}
    \centering
    \includegraphics[width=\linewidth]{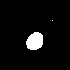}
    \caption{}
  \end{subfigure}
  \begin{subfigure}{0.3\linewidth}
    \centering
    \includegraphics[width=\linewidth]{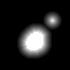}
    \caption{}
  \end{subfigure}
  \caption{(a) A mammographic image patch which includes MCs; (b) The corresponding annotation mask; (c) The corresponding proximity function map with parameters $\xi=10$ and $\alpha=1$.}
  \label{fig:ProximityFunction}
\end{figure}

Our MC segmentation model is formulated as follows: Given a mask generated from reference annotations $M(x,y) \in \{0,1\}$, the MC locations are given by $\{(x_i, y_i)\}$ where $M(x_i,y_i) = 1$. The proximity function is then defined as:
\begin{align}\label{eq:proximity-function}
    P(x,y) &= \max_i g(x,y,x_i,y_i) \\
    \label{eq:transform-func}
    g(x,y,x_i,y_i) &= 
    \begin{cases}
     (e^{\alpha (1-r/\xi)}-1)/(e^\alpha -1), & r \le \xi\\
     0,  & r > \xi
    \end{cases}\\
    r &= \sqrt{(x-x_i)^2+(y-y_i)^2}
\end{align}
where $\alpha, \xi$ are tunable parameters. The function maps MC locations on an exponentially curved surface, expanding to a distance $\xi$ with decay rate $\alpha$ before it vanishes. An example of the transformation is illustrated in Figure \ref{fig:ProximityFunction}. This transformed mask compensated for the fact that we had mixed quality annotations (i.e., point-like and exact) and forced the model to learn information from both the precise locations of MCs and the surrounding background. 

We constructed a model which predicts the proximity function $P(x,y)$ given the image $I(x,y)$. A feature pyramid network (FPN) \cite{Lin2017FPN} was used with  Inception-v4 \cite{Szegedy2017} as the backbone. The FPN architecture was introduced for applications such as region proposal, object detection, and instance segmentation. It adopts a pyramidal shape structure similar to many segmentation networks, such as the U-net \cite{Ronneberger2015}, with an encoder that produces semantic features at different scales and a decoder that combines the encoder features by upsampling them.  

The FPN structure is a suitable architecture because it allows features from all scales to contribute to the final prediction independently. An illustration of the network is presented in Figure \ref{fig:CNN}. The network consists of an encoding and a decoding branch. In the encoding branch, the Inception-v4 architecture was adopted with weights pretrained on ImageNet \cite{Yakubovskiy:2019}. Features were extracted at four different scales (down-sampled compared to the original image by factors of 4, 8, 16, and 32). The features were then transferred to the decoding branch via skip connections. They were upsampled by factors of 1, 2, 4, and 8, respectively, to match their spatial sizes. The resulting features were aggregated using addition and further upsampled to match the image size. The number of output channels was set to 1 and passed through a sigmoid function to generate a value between 0 and 1. This value was thresholded to achieve the final segmentation.

The model was trained using a soft Dice loss function, which was introduced as an optimization objective in biomedical segmentation applications \cite{Milletari2016, Drozdzal2016}. The formulation in \cite{Drozdzal2016} was used:
\begin{align}\label{eq:loss-function}
    L_{\textrm{DICE}}(\hat P, P) = 
    1 - 
     \frac{2 \sum\limits_{x,y} P(x,y) \hat P(x,y) +\epsilon}
     {\sum\limits_{x,y} (P(x,y)+\hat P(x,y)) +\epsilon} 
\end{align}
with $\epsilon$ set to 1 where $\epsilon$ was introduced for numerical stability and $P$ and $\hat P$ correspond to the target and predicted proximity map, respectively. A segmentation binary mask was generated by applying a cut-off on the resulting proximity mask, $\hat P(x,y) \ge p_{thr}$ where $p_{thr} \in [0,1]$.

\subsection{Combining approach}
HDoG resulted in the segmentation of bright objects, whereas the regression CNN outputted a mask of MC regions. These outputs were combined to achieve the  segmentation of individual MCs. The objects identified with HDoG were candidate MC objects, and their shape had to remain intact. Our approach chose a subset of these candidates in their entirety. We retained the HDoG objects that have an overlap $\leq o_{\textrm{thr}}$ with the CNN region mask, where $o_{\textrm{thr}}$ is a tunable percentage parameter.

\subsection{Evaluation metrics}\label{sec:EvaluationMetrics}
The performance of HDoGReg was assessed using Intersection over the Union (IoU). We defined IoU per object as the averaged IoU between each reference annotation object and the object that has the most overlap with in the prediction mask\footnote{MC objects annotated by a single pixel were disregarded in computing IoU per MC object. IoU is not well-defined in such cases.}. To evaluate the image-wise segmentation, the mean IoU between the background and the positive MC class per image was computed. The IoU per MC object was measured to examine the performance of segmenting individual MCs.

HDoGReg was also evaluated as a MC detection task using Free-Response Operating Characteristic (FROC) analysis, similar to prior work \cite{El-Naqa2002, Wang2018}. In FROC analysis, the true positive detection rate was contrasted with false positive detection per image. The analysis required the definition of localization rules to determine true positives. We defined a detected object as a true positive if its distance from a ground truth object was at most 5 pixels (0.35 mm)\footnote{Centroids of individual objects were used in computing their distance.} or if it demonstrated an IoU value of at least 0.3 with a ground truth object.

\section{Experiments}\label{sec:IV}

\begin{figure*}[!h]
  \centering
  \begin{subfigure}{0.2\linewidth}
    \centering
    \includegraphics[width=\linewidth]{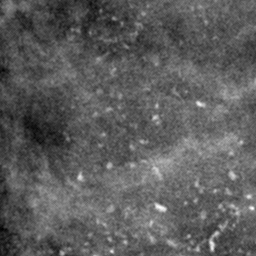}
    \includegraphics[width=\linewidth]{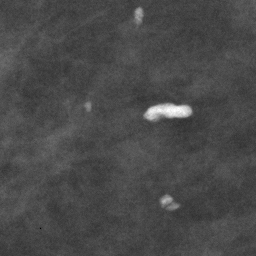}
    \includegraphics[width=\linewidth]{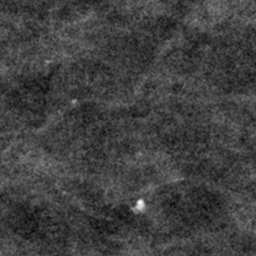}
    \includegraphics[width=\linewidth]{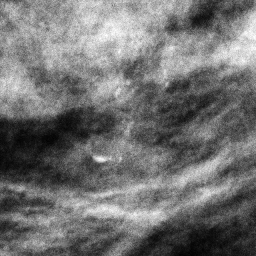}
    \includegraphics[width=\linewidth]{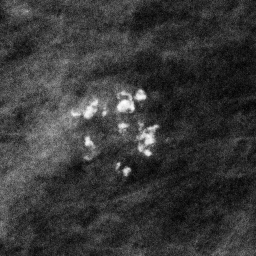}
    \caption{}
  \end{subfigure}
  \begin{subfigure}{0.2\linewidth}
    \centering
    \includegraphics[width=\linewidth]{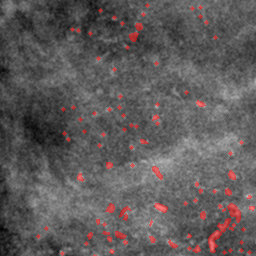}
    \includegraphics[width=\linewidth]{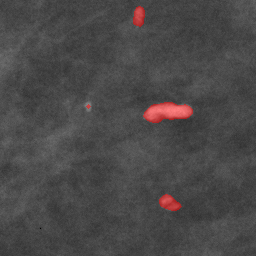}
    \includegraphics[width=\linewidth]{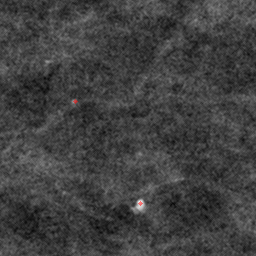}
    \includegraphics[width=\linewidth]{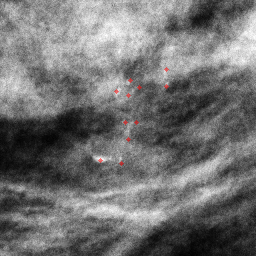}
    \includegraphics[width=\linewidth]{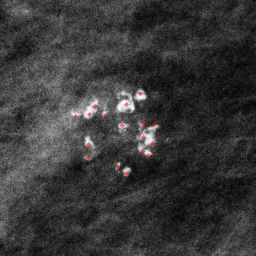}
    \caption{}
  \end{subfigure}
  \begin{subfigure}{0.2\linewidth}
    \centering
    \includegraphics[width=\linewidth]{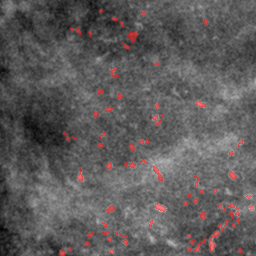}
    \includegraphics[width=\linewidth]{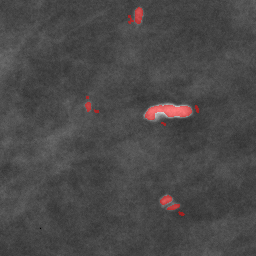}
    \includegraphics[width=\linewidth]{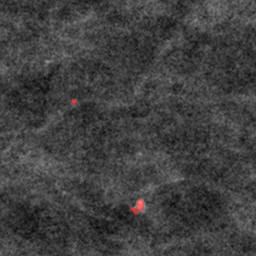}
    \includegraphics[width=\linewidth]{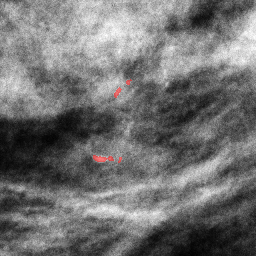}
    \includegraphics[width=\linewidth]{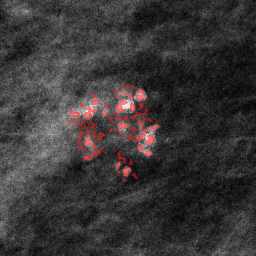}
    \caption{}
  \end{subfigure}
  \begin{subfigure}{0.2\linewidth}
    \centering
    \includegraphics[width=\linewidth]{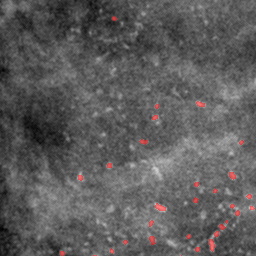}
    \includegraphics[width=\linewidth]{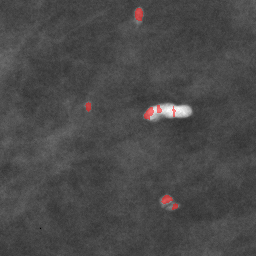}
    \includegraphics[width=\linewidth]{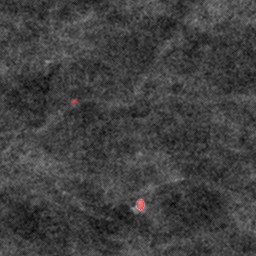}
    \includegraphics[width=\linewidth]{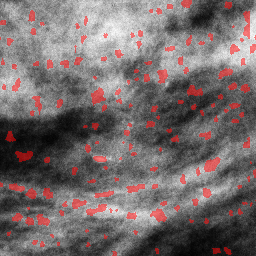}
    \includegraphics[width=\linewidth]{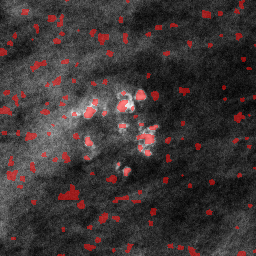}
    \caption{}
  \end{subfigure}
  \caption{Five 256x256 patches extracted from different diagnostic mammograms showing the results of HDoGReg and a comparison method. From left to right: (a) unannotated images, (b) reference annotations, (c) results using HDoGReg, and (d) results using \cite{Ciecholewski2017}. The first three rows are from INbreast data and the last two are from UCLA data. For better visualization the patches were normalized. Note the inherent difference in the appearance of the mammograms between INbreast and UCLA data.}
 \label{fig:SegmentationExamples}
\end{figure*}

\subsection{Training, Validation, and Test Sets}
We partitioned the INbreast dataset into a training set with 51 cases (173 images), a validation set with 17 cases (56 images), and a test set with 18 cases (65  images). The UCLA external test dataset was partitioned into a training set used for fine-tuning the model with 112 images and a held-out test set with 29 images. We used the INbreast validation set to fine-tune HDoGReg, and the INbreast test set to assess the performance. Cases were kept independent (i.e., all images from an individual case were included within the same subset) to avoid potential bias. 

\subsection{Blob Segmentation Optimization}
With HDoG blob segmentation, our goal was to achieve a maximized IoU per object and a true positive rate (TPR) higher than 90\%. We extracted representative patches from the INbreast validation set and optimized HDoG hyperparameters on them to satisfy the above criteria. 

Through experimentation on the validation set, we set $\sigma_{\min} = 1.18$, $\sigma_{\max}=3.1$, overlapping fraction $O_\textrm{DoG}=1$, DoG threshold $T_\textrm{DoG}=0.006$ and Hessian threshold $h_\textrm{ thr}=1.4$. On the validation set, our algorithm achieved an IoU per object 0.648 and a TPR of 0.903 at 136.5 false positives per unit area ($cm^2$).

\subsection{Regression Model Training and Tuning}\label{TrainingSection}
To train the regression model, we extracted patches from the images and corresponding masks. We applied the sliding window approach with patch size 512 pixels and stride 480 to permit overlapping patches. Only patches with annotated MCs present were considered. From INbreast cases, a total of 1045 patches were extracted from the training set and 329 from the validation set. From UCLA cases, a total of 252 patches from the training set were extracted.

The mask patches were transformed using the proximity function map \eqref{eq:proximity-function}. We set $\xi = \{ 6,8,10,12\}$ for the characteristic distance and $\alpha = \{ -1,-2, 10^{-4}, 1,2\}$ for the decay rate. The proximity function \eqref{eq:proximity-function} is not well defined when $\alpha=0$. 

A regression neural network was trained using the image patches of the INbreast training set as input and the corresponding proximity function maps as output. Data augmentation was performed to enrich the training set by randomly applying horizontal flipping, magnification, spatial translations in both directions, cropping, contrast enhancement, brightness adjustment, and gamma correction (details are given in Supplementary Materials). The resulting patches had size 320x320 pixels. The soft Dice loss was used to compute the error between target and predicted proximity functions  \eqref{eq:loss-function}. The model was trained for 40 epochs using the adaptive moment estimation (Adam) optimization method \cite{Kingma2015} with batch size 8, learning rate $10^{-4}$, $\beta_1 = 0.9$, $\beta_2 = 0.999$ and $\epsilon=10^{-8}$. At the end of each epoch, the model was evaluated on the image patches of the INbreast validation set, with the average IoU per patch as the metric. The model achieving the highest IoU over all epochs was kept. The Inception-v4 weights were initialized with weights pretrained on ImageNet. The rest of the model weights were initialized randomly following He initialization \cite{HeInitialization7410480}. The configuration $\xi = 10$, $\alpha=1$ achieved the highest performance for HDoGReg on the validation set. We updated our HDoGReg model to work with data from our institution by updating the model with an additional 40 epochs using a combination of patches from the INbreast and UCLA datasets.

\subsection{Combining Approach Optimization}
The combination approach keeps the HDoG candidate objects based on how much they overlap with the regression model mask. Optimizing the value of $o_\textrm{thr} = \{0.2, 0.3, 0.4, 0.5, 0.6\}$, $o_\textrm{thr}=0.3$ achieved the highest performance on the validation set.

\subsection{Comparison}\label{sec:Comparison}
We compared HDoGReg against two state-of-the-art methods. For the MC detection task, we compared our approach to Wang and Yang \cite{Wang2018}, which used two subnetworks, focusing on the local features and the other on features extracted from the background tissue around the location. They reported a detection performance of 80\% true positive fraction (TPF) at a false positive rate of 1.03 FPs/cm\textsuperscript{2}. We implemented their context-sensitive deep neural network, which classifies a location as MC or non-MC, training our implementation on the INbreast dataset. We implemented DoG based on their reported parameters adjusting the scales to the resolution of our dataset. 

We compared our MC segmentation results with the approach presented in \cite{Ciecholewski2017}. In \cite{Ciecholewski2017}, MC segmentation was performed using morphological operations. In the first step, morphological operators were applied to the original image to detect the MCs' locations. Specifically, a morphological pyramid was generated using  the closing-opening filter. Differences in the pyramid representations of the original image were obtained and combined using the extended maximum of the original image and  morphological reconstruction. In the second step, the MC shapes were extracted using watershed segmentation, where the output of the first step was utilized as a marker. We report the mean IoU per image and the IoU per object for this approach.

\section{Results}\label{sec:V}

\subsection{Regression Model Selection}
The entire segmentation pipeline was evaluated using the INbreast validation set. The FROC analysis is presented in Table \ref{tab:FROCRegModelValidation}, and the mean IoU per image and IoU per object are summarized in Table \ref{tab:SegmRegModelValidation}. For the FROC analysis, a total of 100 bootstrap samples were used to find the partial area under the curve (pAUC) in each experiment. The pAUC was computed for the range between 0 and 1 FPs per unit area, and the 95\% confidence interval is reported. For the computation of the segmentation metrics, a threshold on the predicted proximity function was applied. To determine the optimal threshold for each experiment, we referred to the corresponding FROC curve and found the point closest to TPR 1 and false positives per unit area 0. All configurations performed similarly in terms of the FROC analysis and segmentation metrics. We chose the model with the highest mean value of the  FROC pAUC with $\xi=10$ and $\alpha=1$.

\begin{table}
\caption{Individual MC FROC pAUC values for validating different regression models. The highest pAUC is bolded.}
\centering
\begin{tabular}{l|c|c|c|c}
    \hline
    $\alpha/\xi$            &   6   &   8   &   10  &   12       \\ \hline\hline
    -2                      &   0.804$\pm$0.048   &   0.812$\pm$0.049   &  0.793$\pm$0.040  &   0.773$\pm$0.045   \\ \hline
    -1                      &   0.783$\pm$0.058   &   0.808$\pm$0.047   &   0.790$\pm$0.047  &   0.794$\pm$0.045  \\ \hline
    10\textsuperscript{-4}  &   0.813$\pm$0.054   &   0.783$\pm$0.056   &   0.790$\pm$0.044  &   0.784$\pm$0.048   \\ \hline
    1                      &   0.799$\pm$0.054   &   0.776$\pm$0.056   &   \textbf{0.819$\pm$0.046}  &   0.790$\pm$0.053  \\ \hline
    2                      &   0.775$\pm$0.056   &   0.791$\pm$0.055   &   0.802$\pm$0.049  &   0.789$\pm$0.053   \\ 
    \hline
\end{tabular}
\label{tab:FROCRegModelValidation}
\end{table}

\begin{table}
\caption{Segmentation results of different regression models on validation set. The highest IoUs are bolded.}
\centering
\begin{tabular}{l|c|c|c|c}
    \multicolumn{5}{c}{}\\
    \multicolumn{5}{c}{Mean IoU per image}\\
    \hline
    $\alpha/\xi$            &   6   &   8   &   10  &   12       \\ \hline\hline
    -2	&	0.593$\pm$0.109	&	0.590$\pm$0.102	&	0.576$\pm$0.088	&	0.560$\pm$0.066 \\ \hline
    -1	&	0.590$\pm$0.105	&	0.579$\pm$0.094	&	0.572$\pm$0.089	&	0.566$\pm$0.077 \\ \hline
    10\textsuperscript{-4}	&	0.596$\pm$0.115	&	0.584$\pm$0.098	&	0.586$\pm$0.097	&	0.584$\pm$0.093 \\ \hline
    1	&	\textbf{0.619$\pm$0.125}	&	0.601$\pm$0.109	&	0.583$\pm$0.105	&	0.587$\pm$0.098 \\ \hline
    2	&	0.610$\pm$0.123	&	0.588$\pm$0.108	&	0.592$\pm$0.105	&	0.580$\pm$0.094 \\ \hline
    \multicolumn{5}{c}{}\\
    \multicolumn{5}{c}{IoU per Object}\\
    \hline
    $\alpha/\xi$            &   6   &   8   &   10  &   12       \\ \hline\hline
    -2	&	0.645$\pm$0.207	&	0.645$\pm$0.206	&	0.648$\pm$0.200	&	0.626$\pm$0.228 \\ \hline
    -1	&	0.647$\pm$0.203	&	0.648$\pm$0.201	&	0.643$\pm$0.208	&	0.640$\pm$0.212 \\ \hline
    10\textsuperscript{-4}	&	0.648$\pm$0.201	&	0.644$\pm$0.207	&	0.641$\pm$0.214	&	0.644$\pm$0.207 \\ \hline
    1	&	0.649$\pm$0.200	&	0.646$\pm$0.206	&	0.647$\pm$0.203	&	0.643$\pm$0.208 \\ \hline
    2	&	\textbf{0.650$\pm$0.197}	&	0.648$\pm$0.202	&	0.649$\pm$0.200	&	0.643$\pm$0.208 \\ \hline

\end{tabular}
\label{tab:SegmRegModelValidation}
\end{table}

\subsection{Individual Microcalcifications}

Table \ref{tab:SegmentationResults} reports the segmentation results of HDoGReg on the INbreast validation and test sets. For comparison, the segmentation results of the morphological method of Ciecholewski et al. (see \ref{sec:Comparison}), are also presented. Using the paired Wilcoxon signed-rank test, we showed that HDoGReg achieved superior performance in both mIoU per image and IoU per object for both subsets with $p<0.01$. Figure \ref{fig:SegmentationExamples} presents a sampling of model outputs.

\begin{table}
\caption{Segmentation Results of Final Model on Validation and Test Sets}
\centering
\begin{tabular}{l|c|c}
    \multicolumn{3}{c}{HDoGReg}\\
    \hline
    Metric/dataset     &   Validation  &   Test\\
    \hline\hline
    mean IoU per image  &   0.583$\pm$0.105        &   0.670$\pm$0.121\\
    \hline
    IoU per object      &   0.647$\pm$0.203       &   0.607$\pm$0.250\\
    \hline
    \multicolumn{3}{c}{}\\
    \multicolumn{3}{c}{Ciecholewski \cite{Ciecholewski2017}}\\
    \hline
    Metric/dataset     &   Validation  &   Test\\
    \hline\hline
    mean IoU per image  &   0.517$\pm$0.037       &   0.524$\pm$0.034\\
    \hline
    IoU per object      &   0.408$\pm$0.286       &   0.363$\pm$0.278\\
    \hline
\end{tabular}
\label{tab:SegmentationResults}
\end{table}

\begin{figure}
    \centering
    \includegraphics[width=\linewidth]{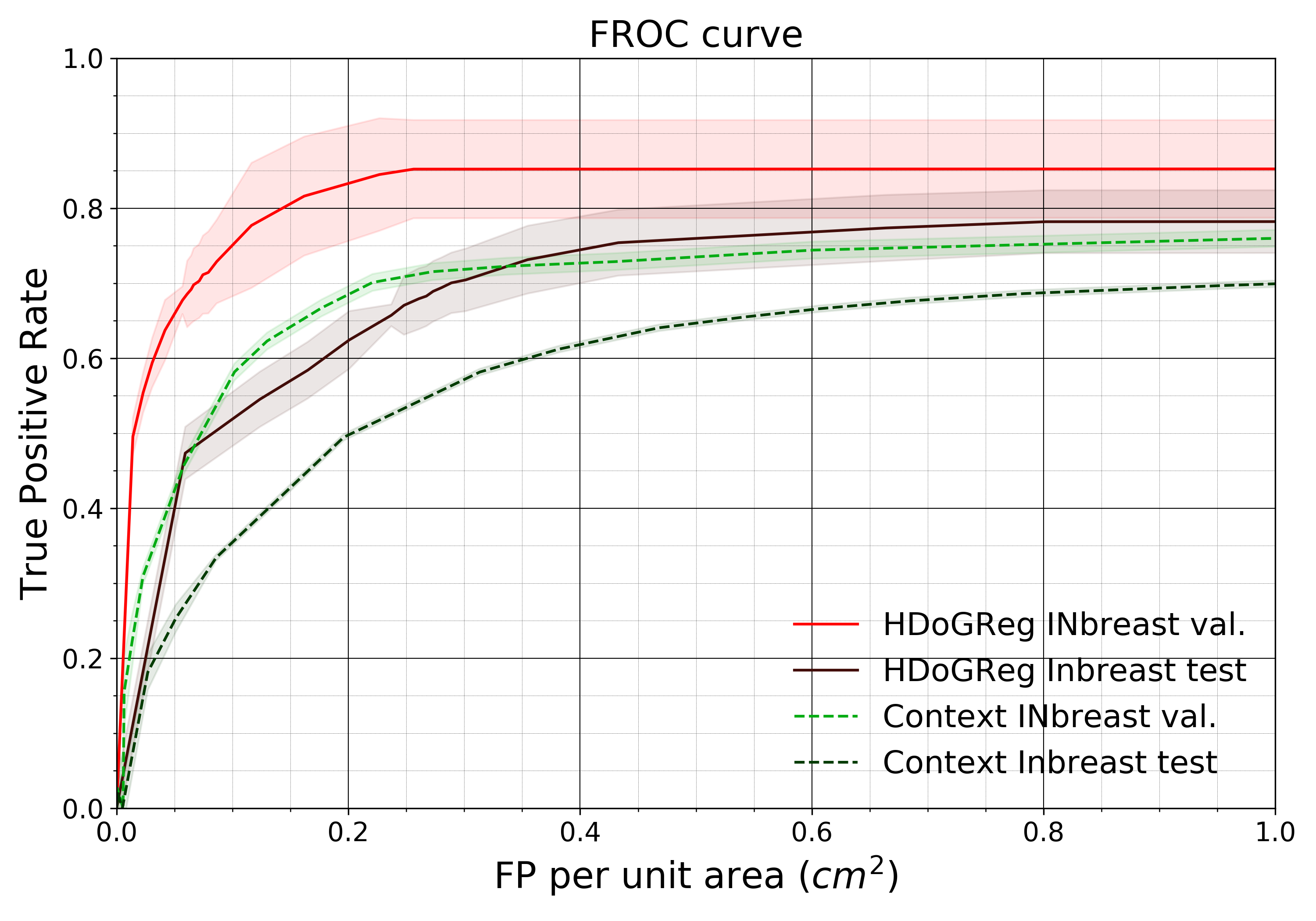}
    \caption{Individual MC FROC analysis for final HDoGReg model and comparison with baseline model.}
    \label{fig:FROCRegModel}
\end{figure}

\begin{figure}
    \centering
    \includegraphics[width=\linewidth]{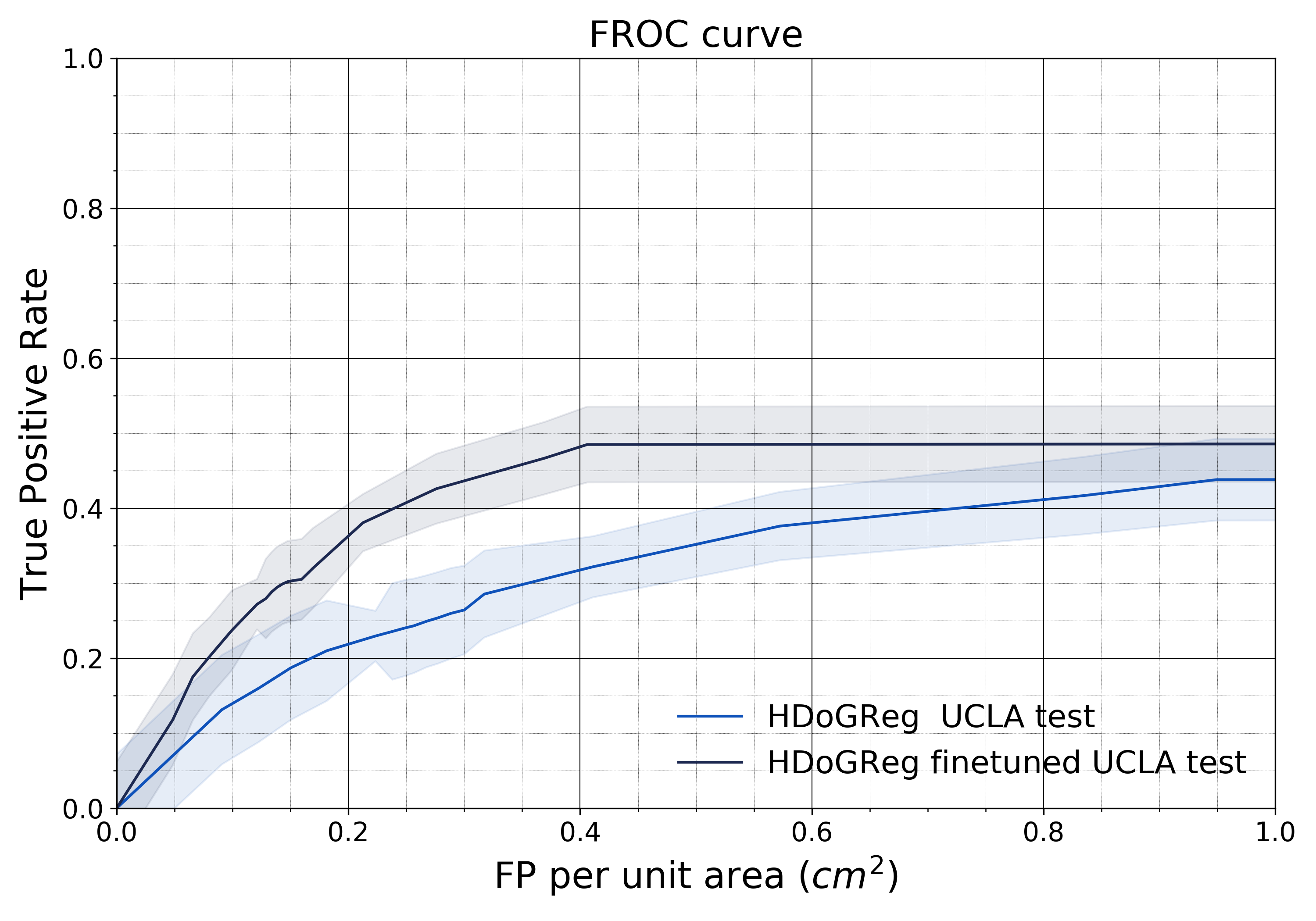}
    \caption{Individual MC FROC analysis on UCLA data.}
    \label{fig:FROCRegModelUCLA}
\end{figure}

Figure \ref{fig:FROCRegModel} presents the individual MC FROC analysis between HDoGReg and \cite{Wang2018}. In this plot, the true positive detection rate on the y-axis was plotted against the false positive counts per unit area (1 $cm^2$). The FROC analysis was performed on the INbreast validation and test sets. HDoGReg achieved FROC pAUC 0.819$\pm$0.046 with a TPR of 0.852 at 0.4 false positives per unit area on the validation set. On the test set, the FROC pAUC was 0.697$\pm$0.078 with a TPR of 0.744 at 0.4 false positives per unit area. In comparison, our implementation of the approach in \cite{Wang2018} achieved FROC pAUC 0.703$\pm$0.057 and  0.581$\pm$0.072 in the validation and test sets.    

 Figure \ref{fig:FROCRegModelUCLA} shows the detection performance of the final model on the UCLA data. We compared the performance of the model trained solely on INbreast data and the model fine-tuned on data from UCLA.  The performance of the two models was comparable since the original model achieved 0.313$\pm$0.109 FROC pAUC and the fine-tuned model achieved  0.420$\pm$0.107. However, for the range 0.2 to 0.6 FPs per unit area, the fine-tuned model outperformed the original based on TPR.

\subsection{Microcalcification Clustering Analysis}
MCs can be clinically significant when they appear in a group or a distribution that would suggest deposits in a duct. In this section, we analyzed HDoGReg segmentation masks to obtain MC clusters and assessed the relationship between the extracted clusters and five predefined groups based on radiologist-interpreted mammographic distributions. To cluster individual MCs into groups, we applied a method called Ordering Points To Identify the Clustering Structure (OPTICS) \cite{Ankerst1999}. OPTICS operates on density-based considerations to identify clusters among a given set of points in a multi-dimensional space. OPTICS defines a notion of neighborhood density of points, based on a threshold for the minimum number of adjacent points within a region of a fixed radius. To perform the clustering, we mapped all MC objects in our segmentation mask to their centroid locations and used them as input to OPTICS. Upon clustering, a total of 24 expert-defined features were extracted describing each cluster's size, shape, and density of MCs (details are given in the supplementary material). The features for all clusters were standardized and fed into K-means clustering with k=5, corresponding to the number of predefined groups. 

The correspondence between the clustering approach versus radiologist descriptors was evaluated as follows: (1) A board-certified fellowship-trained breast radiologist provided labels for mammographic distribution at the image level. Some (n=13, 10.7\%) images were given multiple labels based on the dominant and secondary distribution types that appeared in the image. (2) Each detected cluster was assigned into one of five groups using the k-means algorithm. (3) To associate each point (defined by the features of each cluster) with a distribution label, we duplicated the clusters appearing in images with multiple distribution labels and associated each copy with one distribution label.  Thus, we ended up with a list of distributions and their associated K-means group. (4) We then computed the homogeneity score h \cite{vmeasure2007}, which is a measure describing the purity of all clusters. If all clusters contain points of a single class, h has the value of 1.0. OPTICS was fine-tuned using a random grid search of 100 experiments. We hypothesized that the MC clustering with the highest homogeneity score is the one closest to capturing the actual mammographic distributions. Our clustering approach achieved a homogeneity score $h=0.074$ on the INbreast dataset. Using the extracted quantitative features, we achieved a moderate level of correspondence to predefined radiologist-provided labels of distribution.

\section{Discussion}\label{sec:VI}
We present an approach that combines the difference of Gaussians with Hessian analysis and dense regression to achieve precise MC segmentation in full-field 2D digital mammograms. To our knowledge, this is one of the first works applying a fully convolutional architecture for MC segmentation, which permits concurrent prediction on multiple adjacent locations. The method was trained and validated on 435 mammograms from two separate datasets. The results showed that HDoGReg outperformed comparable approaches that have been recently published. In terms of the FROC analysis using the INbreast dataset, our method achieved a TPR of 0.744 at 0.4 false positives per unit area in comparison with 0.35 TPR at the same false positive counts as what was presented in \cite{Wang2018}. On the segmentation task, HDoGReg achieved 0.670 mean IoU per image and 0.607 IoU per object compared to 0.524 mean IoU per image and 0.363 IoU per object for the morphological approach presented in \cite{Ciecholewski2017}. The addition of UCLA data, even when coarsely annotated by a human reader, improved the performance of HDoGReg. The ability to utilize a mixture of annotations, including exact segmentation of larger calcifications or individual points representing the centroid of smaller calcifications, is a strength of our approach.

While HDoGReg achieved a lower number of false positives compared to other approaches, the overall number of false positives per image is still high. The majority of false positives occur near larger calcifications and correspond to more irregular shapes compared to actual MCs. The irregular detection can be attributed to the regression model which was designed to segment regions containing calcifications. In the case of larger calcifications, the segmented regions span larger areas, increasing the likelihood of retaining false-positive objects. Additional filtering based on size and shape criteria in areas where large calcifications are identified could lead to a substantial false positive reduction. Given that annotating every possible MC is impractical, our algorithm likely identified MCs that were not annotated by human readers, contributing to the false positive count. HDoGReg also  undersegments or oversegments in certain scenarios. Undersegmentation occurs most often in large objects due to: (1) interior regions of objects having lower intensities that are omitted and (2) incorrect delineation of boundaries due to subtle contrast differences between the MC and surrounding tissue. Nevertheless, large calcifications are typically considered benign and not clinically significant. Their undersegmentation will have little effect on quantitative features that may predict invasive cancers. Oversegmentation occurs primarily when bright objects identified with HDoG are close together and erroneously combined into a single object when only part of it corresponds to an actual MC.

Several limitations of our approach exist. Labeling all MCs in full-field mammograms is a time-consuming process and prone to human error and inter-annotator variability. Hence, our work is limited by the dataset size and by variations in how MCs are annotated, ranging from point-like annotations to detailed contours. Our use of a proximity function to reflect the uncertainty associated with MC annotations allows our approach to be robust to training data variations. Moreover, the inherent differences in mammograms acquired with equipment manufactured by different vendors present another challenge. The UCLA dataset was obtained using equipment manufactured by Hologic while the public dataset INbreast were obtained using Siemens equipment. The brightness and contrast levels of the images varied substantially between  manufacturers. Given that the INbreast dataset had four times as many cases as the UCLA dataset, our model was fine-tuned with a limited number of training patches from UCLA. Ongoing work includes annotating additional UCLA cases that would allow us to further fine-tune the model and experiment with different training strategies to improve the generalizability of our approach. Currently, a model with an encoder initialized with pretrained weights was trained first on INbreast data and fine-tuned on mixed UCLA and INbreast data. Provided additional UCLA annotated data, an approach where mixed data are utilized from the beginning of training can be examined. In addition, the significance of using pretrained weights could be explored.  Additionally, magnification views acquired as part of diagnostic mammograms typically improve the conspicuity of confirmed suspicious or benign regions. One direction of future research is to improve the conspicuity of findings in magnification views, potentially yielding more accurate segmentations.

In summary, we presented HDoGReg, a new quantitative approach for MC segmentation, based on blob segmentation and dense regression. We showed that HDoGReg achieves better performance in comparison with state-of-the-art MC segmentation and detection methods. Additionally, we explored how segmented MCs can be clustered in an unsupervised manner, resulting in five groups that are loosely related to qualitative descriptors of MC distributions provided by radiologists. Our results suggest that HDoGReg could be used in large-scale, multi-center studies to define reliable quantitative descriptions of MCs associated with malignancy. Our work serves as the basis for improved quantitative characterization of MCs. A number of shape, intensity, and texture features can be extracted from individually segmented MCs and used to yield quantitative descriptors of MC morphology and distribution. We expect that quantitative descriptors of morphology and distribution may differ from the subjective labels assigned by radiologists. Further studies are needed to evaluate the PPV of our quantitative features and their ability to identify MCs that are associated with invasive cancers.

\section*{Acknowledgment}
 We gratefully acknowledge the support of the NVIDIA Corporation. Also, we would like to thank the Breast Research Group at University of Porto for developing and sharing the INbreast dataset.

\bibliographystyle{IEEEtran.bst}
\bibliography{bibliography}

\end{document}